\documentclass[11pt]{article}
\usepackage[margin=2.5cm]{geometry}
\usepackage{authblk}

\usepackage{amsmath,amssymb}
\usepackage{algorithm}
\usepackage[noend]{algpseudocode}
\usepackage{booktabs}
\usepackage{listings}
\usepackage{xcolor}
\usepackage{graphicx}
\usepackage{url}
\usepackage[colorlinks=true,linkcolor=blue,citecolor=blue,urlcolor=blue]{hyperref}

\newcommand{\qvec}{\mathbf{q}}
\newcommand{\rvec}{\mathbf{r}}
\newcommand{\Fvec}{\mathbf{F}}
\newcommand{\sgmtmp}{\mathsf{B}}
\newcommand{\sgmout}{\sigma}
\newcommand{\code}[1]{\texttt{#1}}
\algrenewcommand\algorithmicrequire{\textbf{Input:}}
\algrenewcommand\algorithmicensure{\textbf{Output:}}
\lstdefinestyle{lmpstyle}{
  basicstyle=\ttfamily\footnotesize,
  keywordstyle=\bfseries,
  commentstyle=\itshape\color{gray},
  morecomment=[l]{\#},
  frame=single, rulecolor=\color{black!30},
  framesep=3pt, xleftmargin=6pt, xrightmargin=2pt,
  breaklines=true, columns=fullflexible, showstringspaces=false,
  morekeywords={compute,fix,run,pair_style,pair_coeff,units,atom_style,
                dimension,read_data,read_restart,neighbor,neigh_modify,
                timestep,thermo,thermo_style,dump,dump_modify,write_restart,
                variable,set,velocity,include}
}

\makeatletter
\@ifundefined{keyword}{%
  \newenvironment{keyword}{\par\noindent\textit{Keywords: }}{\par\medskip}%
}{}
\makeatother
\providecommand{\sep}{, }

\title{kstress: a LAMMPS package for calculating the 2D wavevector-dependent
       stress tensor and its anisotropy in amorphous solids}
\author[1]{O. Benzerara}
\author[1]{J. Baschnagel}
\author[1]{A. Semenov}
\author[2]{L. Klochko$^*$}
\affil[1]{Institut Charles Sadron, CNRS, Universit\'e de Strasbourg, F-67034 Strasbourg, France}
\affil[2]{CRET-AIREL, MosAIk Team, LORIA, Universit\'e de Lorraine,  F-54000 Nancy, France}

\date{}

\begin{document}
\maketitle
\noindent$^*$Corresponding author: liudmyla.klochko@univ-lorraine.fr
\begin{abstract}
The stress tensor resolved in wavevector $\qvec$ and time $t$, $\sigma_{\alpha\beta}(\qvec,t)$,
gives access to the elastic moduli of amorphous solids as a function of the
probed length scale, and to the long-range stress correlations that
distinguish an amorphous solid from a liquid. Its evaluation requires the
contribution of every interacting pair and therefore an interface to the force
loop of the molecular dynamics code. In the current work, we present:
i) an optimized implementation of the LAMMPS~\cite{lammps} module for calculating
$\sigma_{\alpha\beta}(\qvec,t)$, in which trigonometric functions are calculated only for the basic wavevectors  and, in addition,  all the relevant wavevectors are supplied as an
explicit list rather than generated as a full grid, so that only the
wavevectors of interest are computed and stored, and ii) a validation protocol
for the code illustrated for a two-dimensional polydisperse Lennard-Jones (2DpLJ)
glass-forming liquid. Based on these technical achievements we elucidated how the affine moduli for the 2DpLJ model depend on the norm $q$ of the wavevector.  In particular, we demonstrate that the longitudinal affine modulus shows a minimum near $q_\text{max}$  corresponding to the main peak of the structure factor.  Our data also point to the peaks in the  $q$-dependencies of  vibrational frequencies well below  $q_\text{max}$  for both transverse and longitudinal modes.
The potential relevance of this feature to the ``boson peak'' is discussed as well.  
\end{abstract}

\begin{keyword}
stress correlations \sep amorphous solids \sep elastic moduli \sep
molecular dynamics \sep LAMMPS
\end{keyword}

\begin{small}
\noindent{\bf PROGRAM SUMMARY}
\begin{description}\setlength{\itemsep}{1pt}
\item[\it Program Title:] \code{kstress}
\item[\it CPC Library link to program files:] (to be added by Technical Editor)
\item[\it Developer's repository link:] \url{https://github.com/liudmylaklochko/kstress}
\item[\it Licensing provisions:] GPLv2 (as required for LAMMPS-derived source)
\item[\it Programming language:] C++11 (LAMMPS computes), C++11 with OpenMP
      (correlator), Python 3.10 (analysis and validation scripts)
\item[\it External routines/libraries:] LAMMPS (22 July 2025 or later), MPI,
      NumPy
\item[\it Nature of problem:]
  The wavevector-dependent stress tensor $\sigma_{\alpha\beta}(\qvec,t)$ and
  the elastic moduli derived from it are not available in distributed
  molecular dynamics packages. Their evaluation requires the individual
  contribution of every interacting pair, which without a proper optimization
  would drastically slow down simulations.
\item[\it Solution method:]
  The pair contributions are accumulated through the LAMMPS tally callback,
  with no modification of the interaction code. The wavevectors
  are read from a text file rather than generated as a grid, allowing an
  arbitrary set to be computed in one run. Correctness is established by an
  exact identity at $\qvec=0$ against the stress computed internally by LAMMPS,
  and by an independent evaluation over all particle pairs.
\item[\it Additional comments including restrictions and unusual features:]
  The current version is implemented for two-dimensional systems and is not
  applicable to many-body potentials. Bonded interactions are not included. The memory and cost scale with the number
  of wavevectors requested, so a targeted list is preferable to a full grid.
\end{description}
\end{small}

\section{Introduction}
\label{sec:intro}

The mechanical response of an amorphous material is inherently
scale-dependent. In contrast to a crystalline solid, where elastic
properties can be defined with respect to a unit cell and are therefore
naturally associated with macroscopic material constants, an amorphous
solid contains structural heterogeneities over a broad range of length
scales. Local stresses and elastic responses fluctuate from one region to
another~\cite{mizuno2013,nicolas2018}, and the mechanical response obtained
from a measurement depends on the spatial scale over which the material is
probed. A natural description of such systems is therefore provided by
spatially resolved stress and strain fields and, equivalently, by their
Fourier components at finite wavevector $\qvec$, and the natural starting
point is the stress tensor resolved in $\qvec$,
\begin{equation}
\sigma_{\alpha\beta}(\qvec,t)=\frac{1}{V}\int \sigma_{\alpha\beta}(\rvec,t)\,
\mathrm{e}^{-\mathrm{i}\qvec\cdot\rvec}\,\mathrm{d}^d r ,
\label{eq:intro_sigmaq}
\end{equation}
 where $V$ is the volume of the system and $d$ is the space dimension.
Real-space approaches define local elastic moduli on coarse-graining volume
and study their distribution as this volume is varied~\cite{mizuno2013,wittmer2023}. The
result then depends on the size and shape of the volume and on how its
boundary is treated. The reciprocal-space description avoids these issues:
each wavevector probes a well-defined length scale $2\pi/q$ and a
well-defined direction, and the full information is contained in the
correlations of Eq.~\eqref{eq:intro_sigmaq} over a set of wavevectors.

At zero wavevector the connection between stress fluctuations and elasticity
is classical. The shear modulus of a solid can be written as the difference
between an affine contribution, evaluated on single configurations, and a
fluctuation contribution given by the variance of the shear
stress~\cite{lutsko1989,xu2012,wittmer2013}. For truncated pair potentials
the affine modulus must involve an impulsive correction coming from the discontinuity at the
cutoff, which must be included for the decomposition to be
consistent~\cite{xu2012}. The shear relaxation modulus
$G(t)$ follows from the time correlation function of the shear stress through
the fluctuation-dissipation theorem~\cite{hansen2006,allen2017,wittmer2015}.
In glass-forming systems these relations require care: the fluctuation
contribution depends on the statistical ensemble and on the sampling
time~\cite{wittmer2013}, and in a system that does not self-average the
average over independent configurations is itself a fluctuating
quantity~\cite{george2021}. General relations between stress fluctuations
and viscoelasticity, valid for amorphous polymers and glass-forming systems,
have recently been established~\cite{semenov2024}. The question addressed by
the wavevector-resolved description is how these quantities depend on the
length scale over which the deformation is imposed.

Interest in this question has grown from two directions. The first is the
observation that amorphous solids and slowly relaxing liquids carry
\emph{long-range} stress correlations. In a liquid with a short terminal time
the stress field decorrelates over a few particle diameters, whereas in a
glass-forming system it  is not the case:  even upon coarse-graining in time  the stress correlations decay as a power law
$\sim r^{-d}$ in $d$ dimensions, with an angular dependence that survives to
the largest distances the system affords ~\cite{semenov2024,klochko2018,lemaitre2014,lemaitre2015,lemaitre2017,maier2017,maier2018}. Their signature in reciprocal space
is that $\langle|\sigma_{\alpha\beta}(\qvec)|^2\rangle$ retains a dependence
on the \emph{direction} of $\qvec$ down to the smallest accessible
wavevectors.
That anisotropy is not an artefact of a particular coordinate system, nor a
sign of plastic rearrangement: it is imposed by the tensorial structure of an
isotropic medium, and all components of the correlation tensor follow from
the longitudinal and transverse fluctuations
alone~\cite{wittmer2023,wittmer_tensor,wittmer2024}. Establishing it
quantitatively, however, requires the full tensor at many wavevectors, not a
single scalar.

A physical picture at the origin of these correlations is provided by
Eshelby's inclusion problem. A localised rearrangement embedded in an elastic
medium produces a stress perturbation that decays as $r^{-d}$ with a
quadrupolar angular dependence, and the superposition of such events underlies
both the long-range stress correlations of supercooled
liquids~\cite{lemaitre2014,lemaitre2015} and the elastoplastic description of
the plasticity of amorphous solids~\cite{nicolas2018}. In reciprocal space
the corresponding propagator depends on the direction of $\qvec$ but not on
its modulus, which is why the direction dependence of the stress correlations
persists down to the smallest wavevectors. The same structure appears in
theories that derive the long-range correlations from the dynamics of the
supercooled liquid~\cite{maier2017,maier2018}, and in the general analysis of
correlations of tensor fields in isotropic
media~\cite{wittmer2023,wittmer_tensor,wittmer2024}.

The second direction has its roots in generalised hydrodynamics. On molecular
length scales the transport coefficients of a liquid acquire a dependence on
wavevector and frequency~\cite{hansen2006,forster1983,balucani2003,evans2008},
and the wavevector-dependent viscosity kernel has been determined in
simulations of simple and molecular fluids~\cite{puscasu2010}. In the limit
of long wavelengths the transverse and longitudinal correlation functions reduce to the
hydrodynamic description  involving the shear and longitudinal (second)   viscosities,
and the departure from that limit at finite $q$ measures the length scale on
which the continuum description breaks down~\cite{hansen2006,balucani2003}. In supercooled
liquids the viscous response becomes increasingly non-local on approaching
the glass transition~\cite{furukawa2009,furukawa2011}. In liquid theory, the established approach to describe these memory effects is the Zwanzig-Mori projection-operator
formalism~\cite{zwanzig1965,mori1965}. This approach has recently been applied to explore spatio-temporal stress correlations in glass-forming liquids \cite{maier2017,maier2018,vogel2020,grimm2025,PihlajamaaEtal_SPP2023}. In another approach \cite{semenov2024}, the generalised
shear and longitudinal relaxation moduli $G(q,t)$ and $K(q,t)$ were
introduced in~\cite{klochko2018,klochko2022}, and, furthermore, the    relations connecting
them to ordinary $q=0$ correlation functions, the static structure factor and
the thermal conductivity  have been established in  \cite{klochko2022}.   
The important observation is that $K(q,t)$ at $q\neq0$
is essentially independent of the ensemble and of the thermostat, in contrast
to $K(0,t)$. Subsequent work extended these relations to polydisperse
systems,  derived the equilibrium transverse modulus in terms of the generalised
structure factors, and clarified the relationship between the stress noise of
the Newtonian dynamics and the reduced stress of the Zwanzig-Mori
formalism~\cite{grimm2025,sm2026}. 

The connection between stress correlations and
stress memory kernels has been studied for Brownian particles and
viscoelastic fluids~\cite{vogel2020,grimm2025}. At $t=0$ the same correlation
functions give the \emph{instantaneous}, or affine,  moduli related to the stress response
before any rearrangement has occurred~\cite{sm2021,semenov2024}.  What all of this has in common is
that the quantity to be compared with theory is
$\sigma_{\alpha\beta}(\qvec,t)$ which can be measured in simulation over a wide range of
$q$  and over times spanning several decades.

Defining the stress at finite wavevector requires a choice of how the force
between two particles is distributed in space. In the Irving-Kirkwood
construction~\cite{irving1950,tadmor} the contribution of a pair is spread
along the line joining the two particles, which in reciprocal space produces
the factor $\operatorname{e}^{-\mathrm{i}\qvec\cdot\bar\rvec_{ij}}
\operatorname{sinc}(\qvec\cdot\rvec_{ij}/2)$ of Eq.~\eqref{eq:sigmamicro}
below, with $\bar\rvec_{ij}$ being the midpoint of the pair. The per-atom virial
provided by the LAMMPS code~\cite{thompson2009} 
instead assigns half
of each pair contribution to each of the two particles. Summed with the phase
$\operatorname{e}^{-\mathrm{i}\qvec\cdot\rvec_i}$, it replaces the $\operatorname{sinc}$ by
$\cos(\qvec\cdot\rvec_{ij}/2)$. The two definitions coincide at $q=0$ but
differ at order $(\qvec\cdot\rvec_{ij})^2$, a difference that is not
negligible at wavevectors comparable to the inverse particle size. The
per-atom stress is therefore not a substitute for a dedicated calculation.

The ideal contribution to Eq.~\eqref{eq:intro_sigmaq} is a sum over particles
and presents no difficulty, as it can be evaluated from a saved trajectory.
The virial contribution is a sum over interacting pairs, and requires the
individual pair forces. Their calculation must therefore take place inside the force
loop of the molecular dynamics code. Moreover, the phase factor
$\operatorname{e}^{-\mathrm{i}\qvec\cdot\rvec_i}$ attaches to a \emph{particle}, not to a
pair, while the force loop delivers pairs one at a time. No distributed
molecular dynamics package provides the required quantity $\sigma_{\alpha\beta}(\qvec,t)$ and the implementation used
for the results in~\cite{thesis,klochko2022} relied on a modified pair
style, which ties the calculation to one interaction model
(Sec.~\ref{sec:port}). Once $\sigma_{\alpha\beta}(\qvec,t)$ is available, its
time correlation functions must be obtained over several decades in time, which
is done efficiently with multiple-$\tau$
correlators~\cite{ramirez2010,frenkel2002}, and averaged over many
independent configurations, whose preparation at low temperature requires
efficient equilibration methods such as swap Monte
Carlo~\cite{ninarello2017}.

This paper describes an implementation of the calculation of $\sigma_{\alpha\beta}(\qvec,t)$, together with the
analysis chain built around it. Its contributions are as follows:

\emph{(i) An implementation through the standard tally interface.} In
LAMMPS, a pair style reports every interacting pair to the routine
\code{ev\_tally}, which passes it on to every compute registered with the pair
style. We refer to this mechanism as the tally interface, so that
\code{ev\_tally} and the tally interface denote the same thing. Pair
contributions are accumulated in a per-particle buffer and the phase factor
of each particle is applied once, when its contributions are complete
(Sec.~\ref{sec:detail}). Only the standard arguments of the tally interface
are used, so that the compute works with the distributed pair styles,
unmodified, in contrast to the earlier
implementation~\cite{thesis,klochko2022}.

\emph{(ii) An arbitrary set of wavevectors.} The wavevectors are read from a
file rather than generated as a rectangular grid, so that a targeted set may
be requested directly, without recompilation, and only the wavevectors wanted
are computed and stored (Sec.~\ref{sec:qlist}).

\emph{(iii) A validation protocol.} At $\qvec=0$ the
definition reduces to the ordinary stress tensor, so a single row of the
wavevector file provides an exact test against the value the molecular
dynamics code computes internally, at no simulation cost. At finite $q$ we
compare with an independent evaluation over all particle pairs, which shares
no code with the compute. For the ideal contribution the averages can be
written in closed form (Appendix~\ref{app:ideal}), giving a third, analytic
check (Sec.~\ref{sec:valid}).

The implementation is written for LAMMPS~\cite{lammps} and is illustrated on a
two-dimensional polydisperse Lennard-Jones glass former, the system studied
in~\cite{klochko2018,sm2021,klochko2022}, for which independent results
exist to compare. Two-dimensional systems are well suited for this purpose
because, at a given number of particles, the linear size of the box is
larger than in three dimensions so that smaller wavevectors are accessible,
while the extension to three dimensions changes nothing in the construction
described below. Section~\ref{sec:impl} defines the quantities,
Sec.~\ref{sec:analysis} describes the analysis chain, and
Sec.~\ref{sec:valid} gives the validation. Section~\ref{sec:moduli} presents
the wavevector-dependent instantaneous moduli obtained at $t=0$ and relates
them to the relaxation moduli introduced in~\cite{klochko2022}, of which they
are the short-time limit. Section~\ref{sec:discussion} presents a discussion
and summary; the availability of the code is given in the Data Availability
statement. The
algorithm, its cost, the porting to a current version of the molecular
dynamics code and the usage are described in the Appendices
(Secs.~\ref{sec:impl_details}, \ref{sec:port} and \ref{sec:usage}).

\section{Implementation}
\label{sec:impl}
This section defines the quantities computed. The algorithm, the
   wavevector list and the cost are described in
   Appendix~\ref{sec:impl_details}.
\subsection{Definitions}
\label{sec:def}

The microscopic stress tensor at wavevector $\qvec$ is defined, following the
Irving-Kirkwood construction in the form given by Tadmor and
Miller~\cite{tadmor} and by Lema\^itre~\cite{lemaitre2014}, as
\begin{equation}
\sigma_{\alpha\beta}(\qvec,t)
= -\frac{1}{V}\sum_{\langle i,j\rangle}
   F^{\alpha}_{ij}(t)\, r^{\beta}_{ij}(t)\,
   \operatorname{e}^{-\mathrm{i}\qvec\cdot\bar\rvec_{ij}}\,
   \operatorname{sinc}\!\left(\frac{\qvec\cdot\rvec_{ij}}{2}\right)
  -\frac{1}{V}\sum_{j=1}^{N} m_j\, v^{\alpha}_j(t)\, v^{\beta}_j(t)\,
   \operatorname{e}^{-\mathrm{i}\qvec\cdot\rvec_j},
\label{eq:sigmamicro}
\end{equation}
with $\rvec_{ij}=\rvec_i-\rvec_j$, $\bar\rvec_{ij}=(\rvec_i+\rvec_j)/2$,
$F^{\alpha}_{ij}=f_{\rm pair}\,r^{\alpha}_{ij}$, and
$\operatorname{sinc}(x)=\sin x/x$. The first sum runs over distinct pairs
$\langle i,j\rangle$, each counted once, which is how the tally interface of
the MD code delivers them. Instead, summing over ordered pairs $i\neq j$
replaces $\operatorname{sinc}(x)$ by $\operatorname{sinc}(x)/2$, the two forms
being equivalent because $r_{ij}^{\alpha}F_{ij}^{\beta}$ and
$\operatorname{e}^{-\mathrm{i}\qvec\cdot\bar\rvec_{ij}}$ are both invariant under
$i\leftrightarrow j$. Here  $f_{\rm pair}$  is the ratio of the interaction
force between the particles $i,j$ and their distance.  

The phase of a pair is that of its midpoint, so that the contribution is
symmetric in $i$ and $j$. It is convenient to write the virial term as a sum
over particles,
\begin{equation}
-\frac{1}{V}\sum_i \operatorname{e}^{-\mathrm{i}\qvec\cdot\rvec_i}\,
\sgmtmp^{\,i}_{\alpha\beta}(\qvec),
\qquad
\sgmtmp^{\,i}_{\alpha\beta}(\qvec)=\sum_{j\,(\neq i)}
  F^{\alpha}_{ij} r^{\beta}_{ij}\,
  \operatorname{e}^{-\mathrm{i}\qvec\cdot(\bar\rvec_{ij}-\rvec_i)}\,
  \operatorname{sinc}\!\left(\frac{\qvec\cdot\rvec_{ij}}{2}\right),
\label{eq:buffer}
\end{equation}
because this is the order in which the quantity is built: the inner sum
$\sgmtmp^{\,i}$ is accumulated pair by pair as the callback delivers the
neighbours of $i$, and the factor $\operatorname{e}^{-\mathrm{i}\qvec\cdot\rvec_i}$ is
applied to it once, when the accumulation for a given  $i$ is complete. Below, $\sgmtmp$ denotes this
per-particle buffer. At $\qvec=0$ the factor
$\operatorname{e}^{-\mathrm{i}\qvec\cdot\bar\rvec_{ij}}\operatorname{sinc}(0)$ equals unity
and Eq.~\eqref{eq:sigmamicro} reduces to the ordinary stress tensor, which
provides an exact check against the value computed internally by the MD code,
which we use throughout as the primary validation (Sec.~\ref{sec:valid}).

The total stress correlation function is defined as
\begin{equation}
C_{\alpha\beta\gamma\delta}(\qvec,t)=\frac{V}{T}\,
\big\langle \sigma_{\alpha\beta}(\qvec,t_0+t)\,
\sigma^{*}_{\gamma\delta}(\qvec,t_0)\big\rangle ,
\label{eq:Cqt}
\end{equation}
where  $T$ is the temperature (in energy units, i.e.\ Boltzmann constant $k_\text{B} = 1$), $\sigma_{\alpha\beta}(\qvec,t)$ is the total (ideal plus virial) stress
of Eq.~\eqref{eq:sigmamicro}, $\sigma^{*}(\qvec)=\sigma(-\qvec)$, and the
average runs over the gliding time $t_0$ and over independent configurations. 

Eq.~\eqref{eq:Cqt}, and the long-range stress correlations they reveal
in viscoelastic and glass-forming fluids, were introduced in~\cite{klochko2018,klochko2022}.

We now consider two-dimensional systems.
With $\hat{\mathbf e}_1=\qvec/|\qvec|$ and $\hat{\mathbf e}_2=(e_1^y,-e_1^x)$,
the longitudinal and transverse projections are
\begin{equation}
\sigma_{11}=\sum_{\alpha\beta}\sigma_{\alpha\beta}e_1^\alpha e_1^\beta,\qquad
\sigma_{12}=\sum_{\alpha\beta}\sigma_{\alpha\beta}e_1^\alpha e_2^\beta,\qquad
\sigma_{22}=\sum_{\alpha\beta}\sigma_{\alpha\beta}e_2^\alpha e_2^\beta .
\label{eq:proj}
\end{equation}

The wavevectors compatible with the periodic box of side $L$
are $\qvec=(2\pi/L)(n_x,n_y)$ with integer $n_x,n_y$. We use the grid
$n_x,n_y\in\{-n_{q\max},\dots,n_{q\max}\}$ with step $\Delta n=4$ and
$n_{q\max}=120$. Since $\sigma(-\qvec)=\sigma^{*}(\qvec)$, only the half
plane $n_y\ge0$ is computed. A $q$-shell is the set of grid vectors with the
same $n^2\equiv n_x^2+n_y^2$, i.e. the same modulus $q=2\pi n/L$. Shells are
labelled by $(q/q_{\min})^2=n^2$ with $q_{\min}=2\pi/L$, so
$(q/q_{\min})^2=16$ is the smallest wavevector of the grid, $q=8\pi/L$. For
the affine moduli we also average over all shells with $q$ in
$[q-\delta q/2,\,q+\delta q/2]$, $\delta q=0.2$.

Inserting the projections of Eq.~\eqref{eq:proj} into Eq.~\eqref{eq:Cqt} gives the correlation
functions of the projected stress components. In two dimensions four of them
are independent, and we refer to each of them as a \emph{channel}: the shear
channel $C_T\equiv C_{1212}$, the longitudinal channel
$C_\parallel\equiv C_{1111}$, the transverse-normal channel
$C_2\equiv C_{2222}$, and the mixed channel $C_\perp\equiv C_{1122}$.
 
Here the axis~$1$ is along $\qvec$ and defined by the unit vector $\hat{\mathbf e}_1$, while the axis~$2$ is defined by the unit vector $\hat{\mathbf e}_2$ and perpendicular to axis~$1$. Using ``naturally rotated coordinates'' \cite{klochko2018,klochko2022,wittmer2023,wittmer2024}---i.e.\ coordinates expressed in the frame given by axis-$1$ and axis-$2$---%
one arrives at the correlation functions which do not depend on the wavevector direction for isotropic systems (upon averaging over long enough trajectories), so that  the accuracy of these functions can be further enhanced by their pre-averaging over a set of wavevectors  with nearly the same  
$q $ (the $q$-shell).

The $t=0$ values of the four channels are the \emph{affine}, or
instantaneous, moduli. They are obtained from a single configuration and give
the response of the system to an affine deformation in which every particle follows
the imposed strain: the shear modulus $G_A$, the longitudinal modulus $L_A$,
the transverse-normal modulus $N_A$, and the mixed modulus $M_A$ coupling the
two normal components,
\begin{equation}
\begin{aligned}
G_A(q) &= \frac{V}{T}\big\langle|\sigma_{12}|^2\big\rangle, &
L_A(q) &= \frac{V}{T}\big\langle|\sigma_{11}|^2\big\rangle, \\
N_A(q) &= \frac{V}{T}\big\langle|\sigma_{22}|^2\big\rangle, &
M_A(q) &= \frac{V}{T}\big\langle\sigma_{11}\sigma_{22}^{*}\big\rangle .
\end{aligned}
\label{eq:moduli}
\end{equation}

Time dependent stress correlations,  which account for the subsequent rearrangement, follow from
the same correlation functions with one of the $\sigma_{\alpha\beta}(\qvec)$ being replaced by $\sigma_{\alpha\beta}(\qvec,t)$. However,  below we focus on the affine moduli. Note that the time dependent stress correlations were considered earlier 
in~\cite{klochko2018,klochko2022,semenov2024}.

\section{Analysis chain and correlation functions}
\label{sec:analysis}

The output of the computes is a time series of $\sigma_{\alpha\beta}(\qvec)$
on the wavevectors of the list. The correlation functions are formed with a
multiple-$\tau$ scheme~\cite{ramirez2010,frenkel2002}
(Algorithm~\ref{alg:multitau}): the finest level resolves the sampling
interval (with lags $0, \ldots, 2m_s$ at spacing $\delta t$),
 and each further level averages neighbouring samples in pairs and
doubles the lag spacing. A 
 time window $N_f \delta t$ 
 of $N_f$ samples is thus covered with
$\mathcal{O}(m_s\log_2 N_f)$ lags rather than $\mathcal{O}(N_f)$, at constant
relative resolution in $t$, which is what is desired for correlation functions
that decay over several decades. The memory required is that of the series
itself, since no storage grows with the number of lags.

Because all channels are built from the same stored series and on the same
wavevector list, they are obtained in a single pass
(Algorithm~\ref{alg:channels}). The projections~\eqref{eq:proj} are formed
frame by frame, and the correlations of every shell $|\qvec|^2$ are averaged
over the wavevectors it contains.

\begin{algorithm}
\caption{Multiple-$\tau$ correlator~\cite{ramirez2010,frenkel2002}. Level $0$ resolves the sampling interval, and
each further level averages neighbouring samples in pairs and extends the lag
range by a factor of two, so that a time window of $N_f\,\delta t$ is covered
with $\mathcal{O}(m_s\log_2 N_f)$ lags instead of $\mathcal{O}(N_f)$, at
constant relative resolution.}
\label{alg:multitau}
\begin{algorithmic}[1]
\Require complex series $A_n, B_n$, $n=0,\dots,N_f-1$, sampled every
         $\delta t$, block size $m_s$
\Ensure  $C(t)=\langle A(0)B^{*}(t)\rangle$ on the multiple-$\tau$ lag set

\State $k_{\max} \gets \lfloor \log_2 (N_f/2m_s) \rfloor$, \quad $N_0 \gets N_f$
\Statex
\State \textbf{level 0:} \Comment{lags $0\ldots 2m_s$ at spacing $\delta t$}
\For{$n = 0$ \textbf{to} $2m_s$}
   \State $C_0(n) \gets \dfrac{1}{N_0-n}\sum_{i=0}^{N_0-n-1} A_i\,B^{*}_{i+n}$
\EndFor
\Statex
\For{$k = 1$ \textbf{to} $k_{\max}$}   \Comment{coarse-grain and extend}
   \State $A_i \gets \tfrac{1}{2}(A_{2i}+A_{2i+1})$, \quad
          $B_i \gets \tfrac{1}{2}(B_{2i}+B_{2i+1})$, \quad $N_k \gets N_{k-1}/2$
   \For{$n = m_s+1$ \textbf{to} $2m_s$}
      \Comment{only the new lags, since $n\le m_s$ is already covered by level $k-1$}
      \State $C_k(n) \gets \dfrac{1}{N_k-n}\sum_{i=0}^{N_k-n-1} A_i\,B^{*}_{i+n}$
      \State lag time $t_{k,n} = n\,2^{k}\,\delta t$
   \EndFor
\EndFor
\Statex
\State \textbf{shell average:} for each $|\qvec|^2$, average $C$ over the
       wavevectors of that shell stored in the file
\end{algorithmic}
\end{algorithm}

\begin{algorithm}[t!]
\caption{Stress channels formed from the compute outputs. At $t=0$ they give
the moduli of Eq.~\eqref{eq:moduli}.}
\label{alg:channels}
\begin{algorithmic}[1]
\Require per-frame arrays on a common wavevector list: the virial and ideal
         parts of $\sigma_{\alpha\beta}(\qvec)$
\Ensure  time correlation functions per $|\qvec|^2$ shell

\State $\sigma_{\alpha\beta} \gets \sigma^{\rm vir}_{\alpha\beta}
        + \sigma^{\rm id}_{\alpha\beta}$
\State project onto $\hat{\mathbf e}_1,\hat{\mathbf e}_2$
       (Eq.~\eqref{eq:proj}) to obtain $\sigma_{11},\sigma_{12},\sigma_{22}$
\Statex
\State \textbf{stress:} $C_{1212},\ C_{1111},\ C_{2222},\ C_{1122}$
       \Comment{$\to G_A, L_A, N_A, M_A$ at $t=0$}
\State \quad optionally the same four for the ideal and virial parts separately
\Statex
\For{each channel}
   \State apply Algorithm~\ref{alg:multitau}, then average over the
          wavevectors of each shell
\EndFor
\end{algorithmic}
\end{algorithm}

\section{Validation}
\label{sec:valid}

The implementation is checked at four levels, of which the first is exact and
should be run first in any new system.

\paragraph{(i) The $\qvec=0$ identity.}
Including the row $(0,0)$ in the wavevector list reduces
Eq.~\eqref{eq:sigmamicro} to the ordinary stress tensor. Comparing it with
the value computed internally by the MD code, split into its virial and
kinetic parts, tests the pair loop, the virial products, the completion of
each particle, and the parallel reduction, in a single number per component.
In the test system used here the six components agree with
\code{compute pressure} to machine precision, separately for the virial and
the kinetic contributions.

\paragraph{(ii) An independent all-pairs calculation.}
A reference implementation in Python evaluates Eq.~\eqref{eq:sigmamicro}
directly over all $N(N-1)/2$ pairs with the minimum image convention, from a
saved trajectory and without any part of the MD code. Over the wavevectors
tested the largest deviation from the compute is $10^{-13}$, which is the
precision of the trajectory dump.

\begin{figure}[t!]
    \centering
    \includegraphics[width=0.48\linewidth]{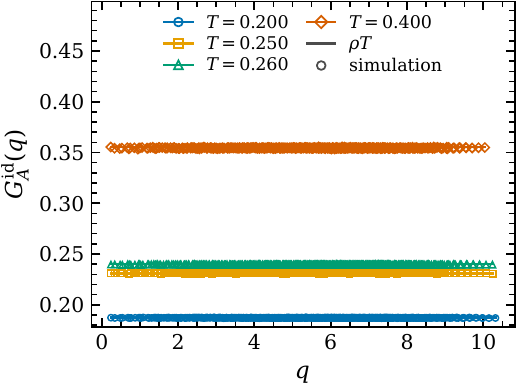}
    \includegraphics[width=0.48\linewidth]{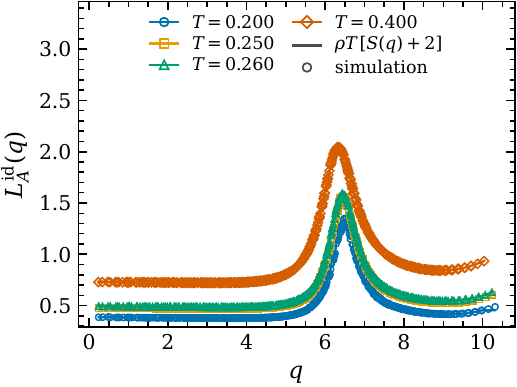}\\
    \includegraphics[width=0.48\linewidth]{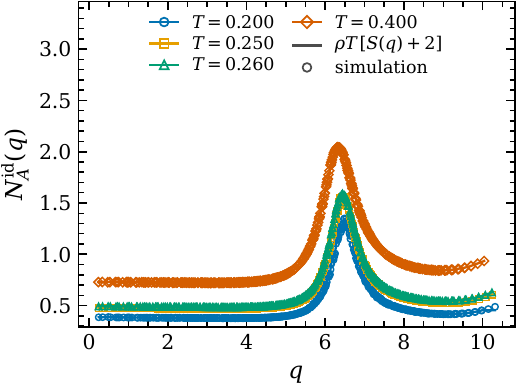}
    \includegraphics[width=0.48\linewidth]{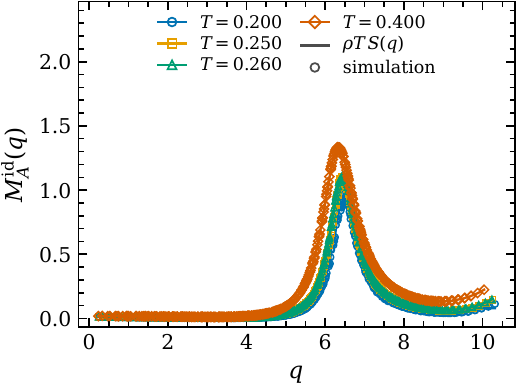}
    \caption{Analytical evaluation of the kinetic  (ideal)     contributions to the wavevector-dependent instantaneous (affine) moduli. From top left to bottom
    right: $G_A^{\rm id}(q)$, $L_A^{\rm id}(q)$, $N_A^{\rm id}(q)$, and $M_A^{\rm id}(q)$.}
    \label{fig:affine_moduli_id}
\end{figure}

\paragraph{(iii) The ideal part in closed form.}
The kinetic contribution can be evaluated analytically,
\begin{equation}
G_A^{\rm id}=\rho T,\qquad
L_A^{\rm id}=N_A^{\rm id}=\rho T\,[S(q)+2],\qquad
M_A^{\rm id}=\rho T\,S(q),
\label{eq:idealcheck}
\end{equation}
with $S(q)$ the static structure factor, which we obtain on the same
wavevector list. The $+2$ arises from the Gaussian fourth moment
$\langle u^4\rangle=3\langle u^2\rangle^2$ of a single velocity component
together with the removal of the self term in
$\sum_{j\neq k}=N S(q)-N$, and is absent from $G_A^{\rm id}$ and
$M_A^{\rm id}$, for which the two relevant velocity components     ($u,w$) refer to perpendicular directions, hence 
$\langle u^2w^2\rangle=\langle u^2\rangle\langle w^2\rangle$
(Appendix~\ref{app:ideal}). This tests the kinetic compute and the projections
independently of the pair loop,  see Fig.~\ref{fig:affine_moduli_id}.

It is interesting to note that kinetic contributions to the affine moduli $L_A^{\rm id}$,  $M_A^{\rm id}$, $N_A^{\rm id}$  show a peak at $q=q_\mathrm{max} \approx  2\pi$  (in LJ units), while the total affine modulus $L_A(q)$ exhibits a minimum there 
(cf.~Fig.~\ref{fig:affine_moduli}).

\paragraph{(iv) Invariances.}
The output is bit-identical to that of the earlier
code~\cite{thesis,klochko2022} when the list reproduces its grid. It is unchanged 
to round-off, under a change in the number of MPI ranks, and unchanged under a
change of neighbor-list skin when the same trajectory is re-analyzed. For 
projections, $\sigma_{11}$ reduces exactly to $\sigma_{xx}$ or $\sigma_{yy}$
when $\qvec$ lies along the coordinate axis $x$ or $y$, which verifies the construction of
the transverse direction.

\subsection{Structural validation}
\label{sec:structural_validation}

\begin{figure}[t]
    \centering
    \includegraphics[width=0.48\linewidth]{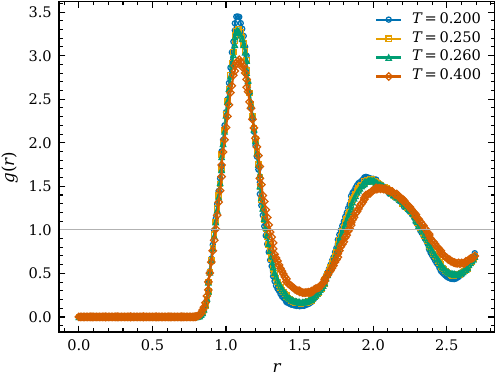}
    \includegraphics[width=0.48\linewidth]{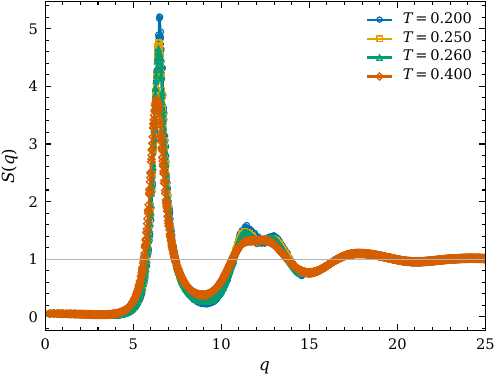}
    \caption{Structural characterization of the simulated 2D
    polydisperse Lennard-Jones system. Left: radial distribution
    function $g(r)$. Right: static structure factor $S(q)$.}
    \label{fig:structure}
\end{figure}

Before analysing the wavevector-dependent stress, we first verify that the
simulated configurations reproduce the expected structural properties of
the polydisperse Lennard-Jones system. Fig.~\ref{fig:structure} shows
the radial distribution function $g(r)$ and the corresponding static
structure factor  $S(q)$ \cite{hansen2006}, both of which were analyzed for the 2DpLJ model before \cite{sm2021}.

The short-range structure and the absence of
long-range crystalline order are consistent with the amorphous character
of the simulated system. The same wavevector list used for the
wavevector-dependent stress calculation is used to evaluate $S(q)$, see  Fig.~\ref{fig:structure}b.

\section{Instantaneous moduli and relaxation moduli}
\label{sec:moduli}

\begin{figure}[t]
    \centering
    \includegraphics[width=0.8\linewidth]{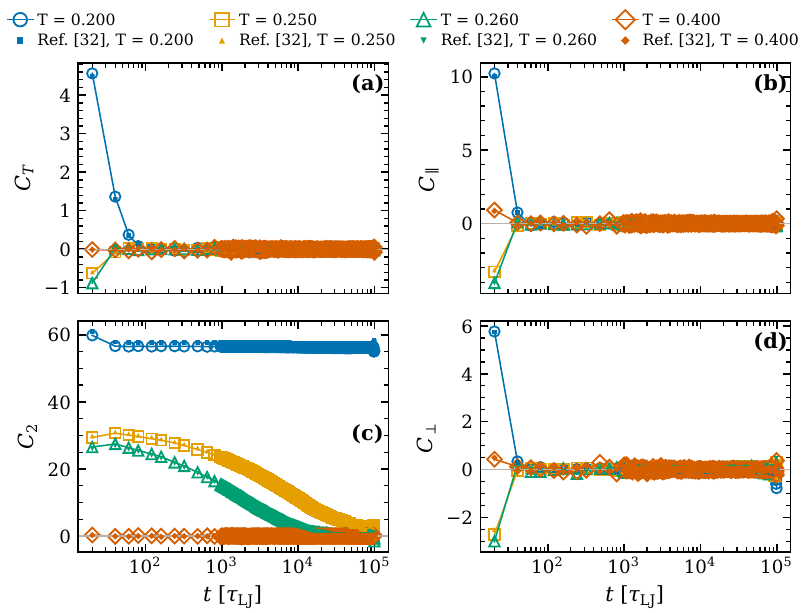}
    \caption{Stress correlation functions $C_T$, $C_\parallel$, $C_2$ and
    $C_\perp$ at the four temperatures, at the wavevector,
    $(q/q_{\min})^2=16$, compared with the data of Ref.~\cite{klochko2022}.}
    \label{fig:corr}
\end{figure}

\begin{figure}[t]
    \centering
    \includegraphics[width=0.8\linewidth]{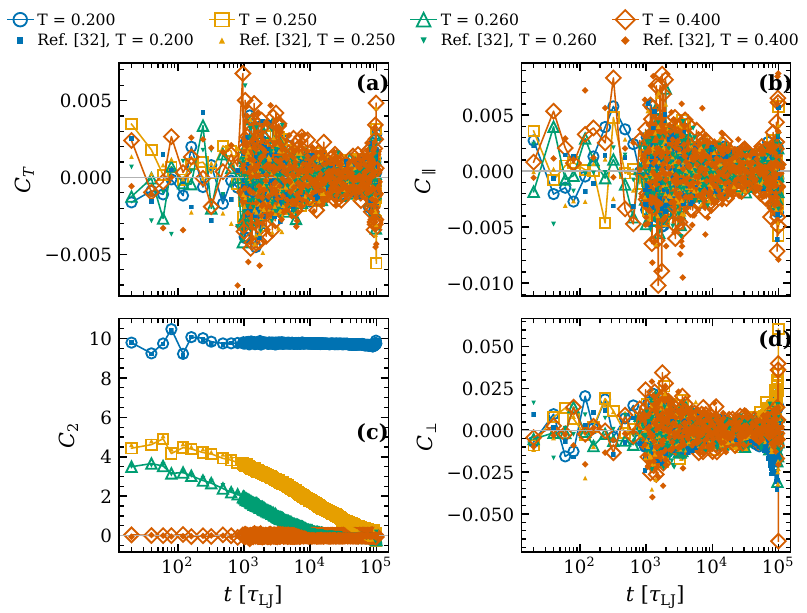}
    \caption{Stress correlation functions $C_T$, $C_\parallel$, $C_2$ and
    $C_\perp$ at the four temperatures, at the highest wavevector,
    $(q/q_{\min})^2=28800$, compared with the data of Ref.~\cite{klochko2022}.}
    \label{fig:corr_1}
\end{figure}

The compute delivers $\sigma_{\alpha\beta}(\qvec,t)$, and the analysis chain
turns it into the four channels defined in Sec.~\ref{sec:def}. It is worth
stating precisely how the instantaneous stress correlation functions (affine elastic moduli) of Eq.~\eqref{eq:moduli}
behave in relation to the relaxation of these correlations considered in~\cite{klochko2022}.
 Figures~\ref{fig:corr} and~\ref{fig:corr_1}
show the most interesting examples of transverse stress correlations, $C_2(q,t)=C_{2222}(q,t)$, see Eqs.~\eqref{eq:Cqt} and \eqref{eq:proj}, for the smallest and largest wavevectors of the grid: $(q/q_{\min})^2=16$ and $(q/q_{\min})^2=28800$. 

\subsection{Relation to the relaxation moduli}
\label{sec:moduli_discussion}

The four quantities of Eq.~\eqref{eq:moduli} are \emph{instantaneous}, or
affine, moduli defined by the $t=0$ values of the corresponding stress correlation functions.
Each affine modulus is a property of a single configuration, which is eventually averaged over production trajectories.   They correspond to the stress increment that
appears immediately upon a deformation in which every particle is displaced
according to the imposed strain, before any rearrangement has taken place \cite{klochko2022}.

The affine moduli are related to, but must be distinguished from, the generalized
relaxation moduli $G(q,t)$ and $K(q,t)$ introduced in~\cite{klochko2022}, which are
defined via the stress response following a canonical affine deformation
applied at $t=0$ and which decay with time.  The instantaneous
moduli are therefore the $t\to0$  limits of relaxation moduli,
\begin{equation}
G_A(q)=\lim_{t\to 0}G(q,t)\big|_{\rm affine},\qquad
L_A(q)=\lim_{t\to 0}K(q,t)\big|_{\rm affine},
\label{eq:affine_vs_relax}
\end{equation}
while the long-time limits give the static moduli. 

\subsection{Wavevector dependence}
\label{sec:qdep}

It is worth highlighting two features of the $q$-dependence that are made accessible via the implementation described here.
\paragraph{The channels do not coincide.}
In an isotropic medium the tensor $\sigma_{\alpha\beta}(\qvec)$ is completely
characterised by its projections onto $\hat{\mathbf e}_1=\qvec/|\qvec|$ and the
perpendicular direction, and the four correlations of Eq.~\eqref{eq:moduli}
are not independent functions of $|\qvec|$ alone: their differences at fixed
$|\qvec|$ encode the direction dependence of the underlying real-space
correlations. An instantaneous stress field correlated only over a few particle diameters
would give channels that converge as $q\to0$, whereas the persistence of a splitting
down to the smallest accessible wavevector is the reciprocal-space signature
of the long-range correlations discussed
in~\cite{klochko2018,lemaitre2014,wittmer_tensor}. It is this splitting, and
not any individual curve, that carries the physical content, which is why the
implementation stores the Cartesian components and forms all four channels
from the same data.

\begin{figure}[t]
    \centering
    \includegraphics[width=0.48\linewidth]{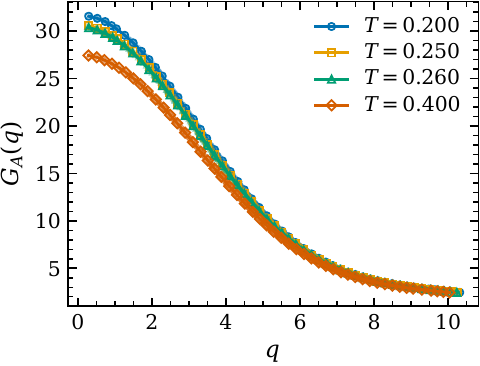}
    \includegraphics[width=0.48\linewidth]{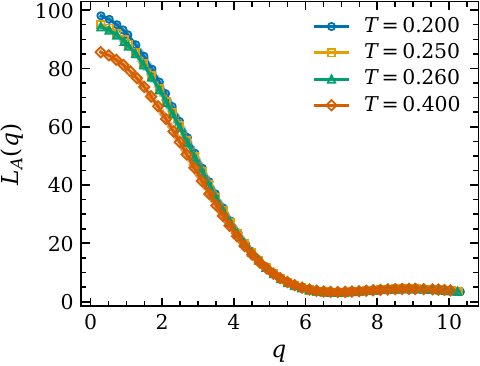}\\
    \includegraphics[width=0.48\linewidth]{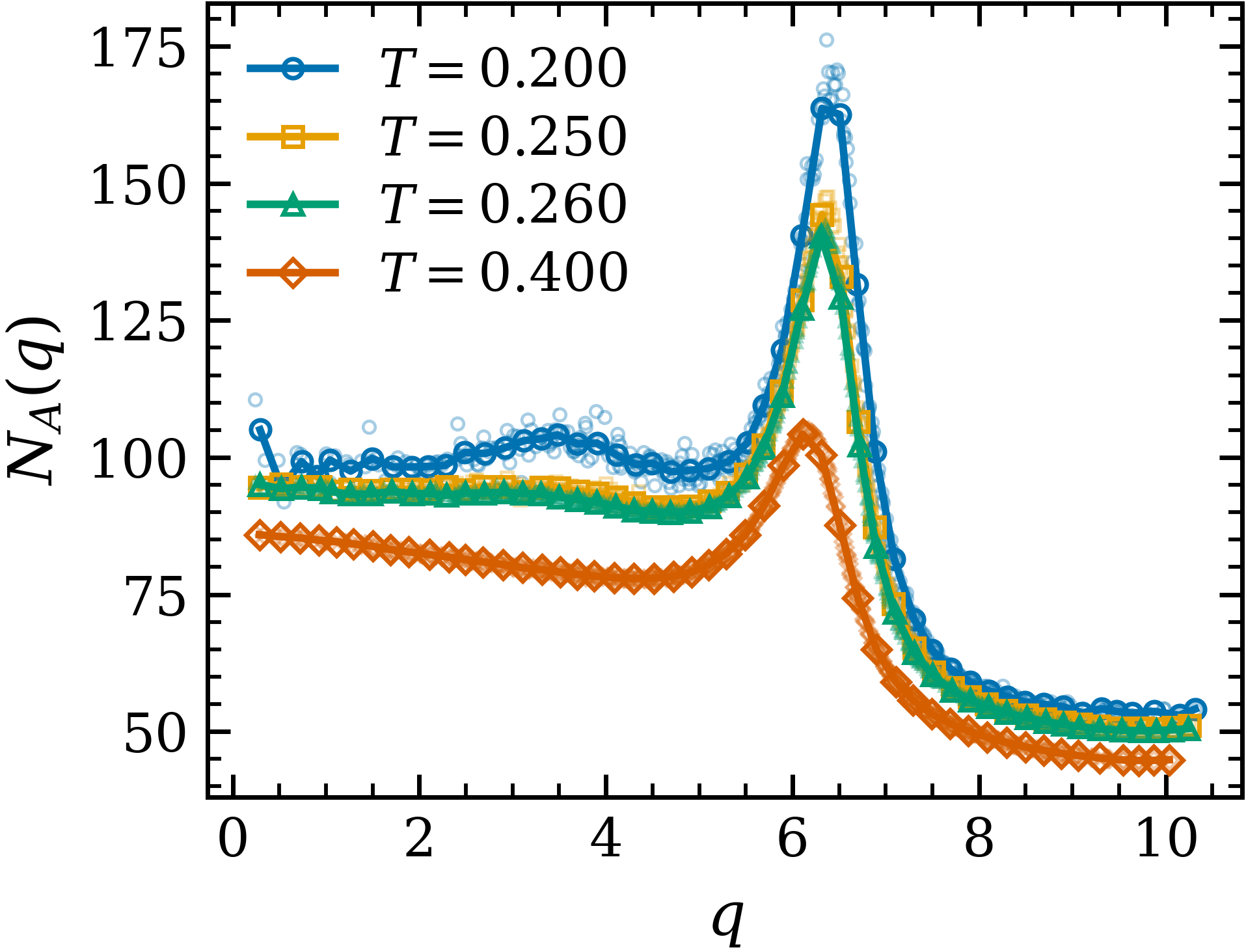}
    \includegraphics[width=0.48\linewidth]{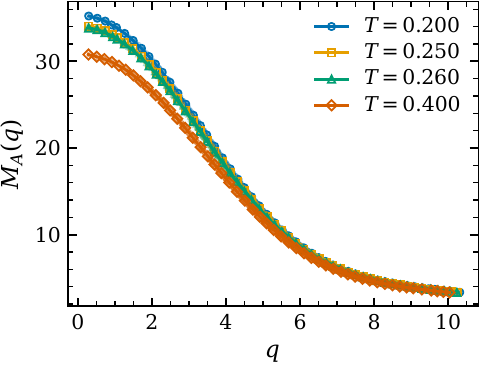}
    \caption{Wavevector-dependent instantaneous (affine) moduli obtained
    from the zero-time stress correlations. From top left to bottom
    right: $G_A(q)$, $L_A(q)$, $N_A(q)$, and $M_A(q)$.}
    \label{fig:affine_moduli}
\end{figure}

The resulting wavevector dependence of the four instantaneous moduli is
shown in Fig.~\ref{fig:affine_moduli}. All four quantities are obtained
from the same microscopic stress tensor, but correspond to different
tensorial projections of the stress fluctuations. The shear modulus
$G_A(q)$ characterizes the shear stress response, while $L_A(q)$, $N_A(q)$ and $M_A(q)$ are related to the longitudinal, normal and transverse components of the stress correlation tensor (see Fig.~\ref{fig:affine_moduli}).

\begin{figure}[t!]
    \centering
    \includegraphics[width=1.\linewidth]{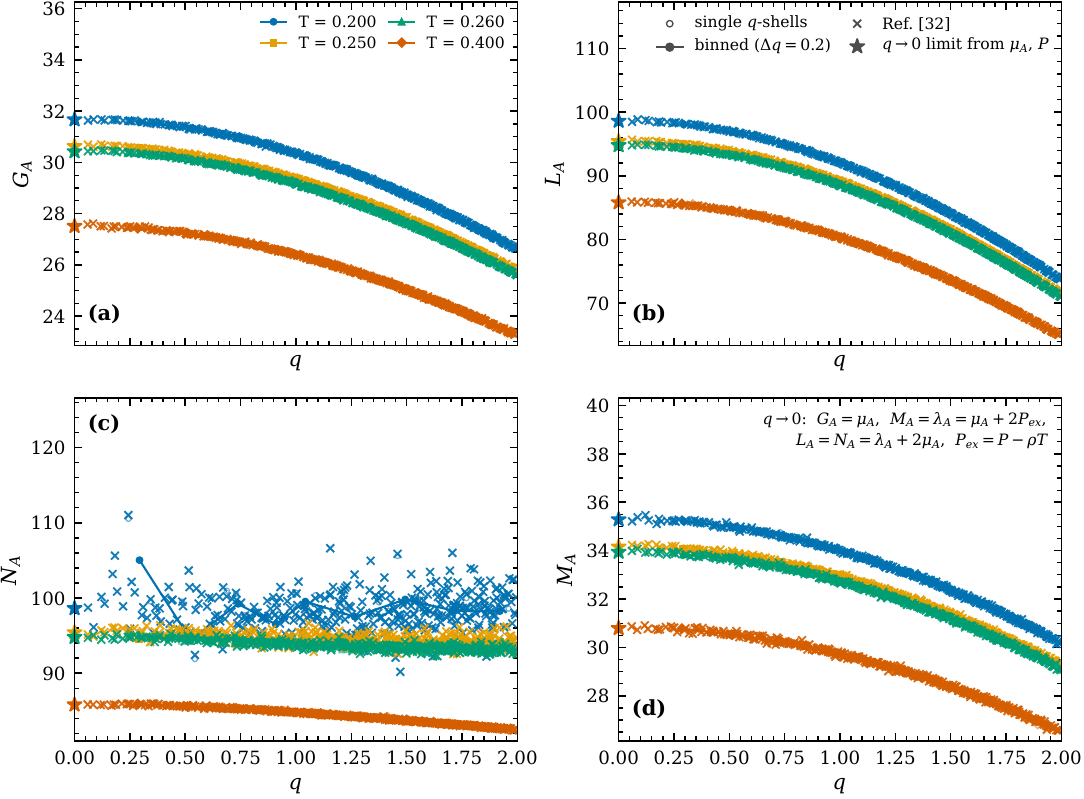}
    \caption{Affine moduli (a) $G_A$, (b) $L_A$, (c) $N_A$ and (d) $M_A$
    for $q<2$ at the four temperatures. Open symbols: single $q$-shells;
    filled symbols with lines: average over $\delta q=0.2$; crosses:
    Ref.~\cite{klochko2022}; stars: $q\to0$ limits from $\mu_A$ and $P$,
    Eq.~\eqref{eq:q0}.}
    \label{fig:newFigRef29}
\end{figure}

For $0<q<2$, Fig.~\ref{fig:newFigRef29} compares the affine moduli with the
data of Ref.~\cite{klochko2022}, obtained from shorter trajectories on a
coarser wavevector grid. The two data sets agree within their scatter.
$N_A$ scatters more than the other moduli at $T=0.2$. This scatter comes
from the frozen part of $\sigma_{22}$, which below $T_g$ does not average
out over a single trajectory, so it depends on the configuration.

\paragraph{The ideal contribution is known exactly.}
Because the velocities are Maxwellian and independent of the positions, the
kinetic part of each modulus is given in closed form by
Eq.~\eqref{eq:idealcheck} (Appendix~\ref{app:ideal}). Separating the ideal and
virial contributions therefore yields an internal consistency check at every
wavevector at no extra cost, since $S(q)$ is accurately computed on the same wavevector
list. In practice, this is the most sensitive of the routine tests: a
normalisation error, a misassigned column or a wrong dimension shows up
immediately as a constant-factor discrepancy in Eq.~\eqref{eq:idealcheck},
whereas it might pass unnoticed in the total moduli.

\subsection{The $q\to0$ limit}

At $q\to0$ the instantaneous moduli must reduce to those of a uniform
deformation: $G_A(q\to0)=\mu_A$, and the combinations of $L_A$, $N_A$ and
$M_A$ reduce to the affine bulk and shear moduli \cite{semenov2024}. Since $\mu_A$ is available
independently from the standard uniform-shear expression~\cite{lutsko1989},
this limit is a check on the $q$-resolved data that does not involve the
compute at all. A similar verification is also possible for the other 3 moduli.  
Here the important point is that the affine moduli are related to instantaneous stress correlations (with no time-shift), which are always short-range in well-equilibrated systems.     

For a pair potential in two dimensions the $q\to0$ limits follow from $\mu_A$
and the pressure alone~\cite{zwanzig_mountain1965,lutsko1989}:
\begin{equation}
G_A(0)=\mu_A,\qquad
M_A(0)=\lambda_A=\mu_A+2P_{\rm ex},\qquad
L_A(0)=N_A(0)=\lambda_A+2\mu_A,
\label{eq:q0}
\end{equation}
with $P_{\rm ex}=P-\rho T$ the excess pressure. These values, computed
without the stress compute, are shown as stars in
Fig.~\ref{fig:newFigRef29}. All four moduli reach well-defined plateaus at
low $q$ (Fig.~\ref{fig:affine_moduli}), and the plateaus agree with
Eq.~\eqref{eq:q0} at all temperatures.

By contrast, below the glass transition, the  simple relations between the 4 moduli  \cite{semenov2024} get strongly violated for any $q$ at long time-shifts  $t$ \cite{klochko2022} : only $C_{2222}(q,t)$ (corresponding to $N_A$ at $t=0$) survives at long  $t$  beyond the sound attenuation time.  This behavior leads to a strong dependence of the stress correlation tensor on the direction of $\qvec$  even at low q's  \cite{klochko2018,wittmer2023} thereby providing the reciprocal-space signature of the
long-range stress correlations discussed in~\cite{klochko2018,klochko2022}.
The tensorial structure that governs this behavior is described
in~\cite{wittmer_tensor,wittmer2023,wittmer2024}. The extrapolation to
$q\to0$ is therefore not always trivial, which is one of the reason why resorting to the
smallest wavevector $q_{\min}$, which the system size allows, can be useful.

\subsection{Ideal and virial parts}

Equation~\eqref{eq:sigmamicro} decomposes into a kinetic and a virial term
and the compute delivers them separately, so the $t=0$ correlations give the
ideal and virial contributions to each modulus. The ideal contribution is
known in closed form (Appendix~\ref{app:ideal}) and is small at the
temperatures of interest, of order $\rho T$ against instantaneous moduli
of order $10$ to $10^2$, but its exact form is a test in its own right, and
the separation is needed for a clean comparison with theory, since the
relations derived in~\cite{klochko2022} treat the two parts differently at
finite $q$. One point deserves emphasis: the total correlation is
$\langle|\sigma^{\rm id}+\sigma^{\rm vir}|^2\rangle$, not the sum of the two
autocorrelations, and the cross term
$2\,\mathrm{Re}\langle\sigma^{\rm id}\sigma^{\rm vir*}\rangle$ is not
negligible relative to the ideal part. A naive decomposition that just adds the two
autocorrelations is therefore not sufficient.

\subsection{Below the glass transition}

Above $T_g$ all four correlation functions decay to zero and the
instantaneous moduli can serve as the    main benchmark. Below $T_g$   the shear
stress at $q=0$  acquires a frozen component so that $\langle\sigma_{xy}\rangle$ averaged over the system volume no longer vanishes, and the shear-stress correlation function decay to 
a finite plateau~\cite{semenov2024,klochko2022}.   
By contrast, at  $q>0$  and even well below $T_g$  the three correlation functions $C_{1212}=C_{2121}$, $C_{1111}$ and $C_{2211}=C_{1122}$ decay to zero at long times, while $C_{2222}(q,t)$ attains a finite plateau value defined by the frozen-in (quenched) stress  \cite{klochko2022}.

\section{Discussion and summary}
\label{sec:discussion}

1. Discussion on the data for $q$-dependent affine moduli

We considered 4 types of affine moduli:  shear $G_A(q)=G(q,t=0)$, 
longitudinal $L_A(q)=L(q,t=0)$, mixed (transverse) $M_A(q)=M(q,t=0)$, 
and normal (perpendicular) $N_A(q)=N(q,t=0)$. The data are obtained 
for a wide range of wavenumbers $q$ (see Fig.~\ref{fig:affine_moduli}); they are verified against the 
previously obtained data~\cite{klochko2022} available for a much narrower range 
of $q$, $0<q<2$, see Fig.~\ref{fig:newFigRef29}.

\begin{figure}[t]
    \centering
    \includegraphics[width=0.6\linewidth]{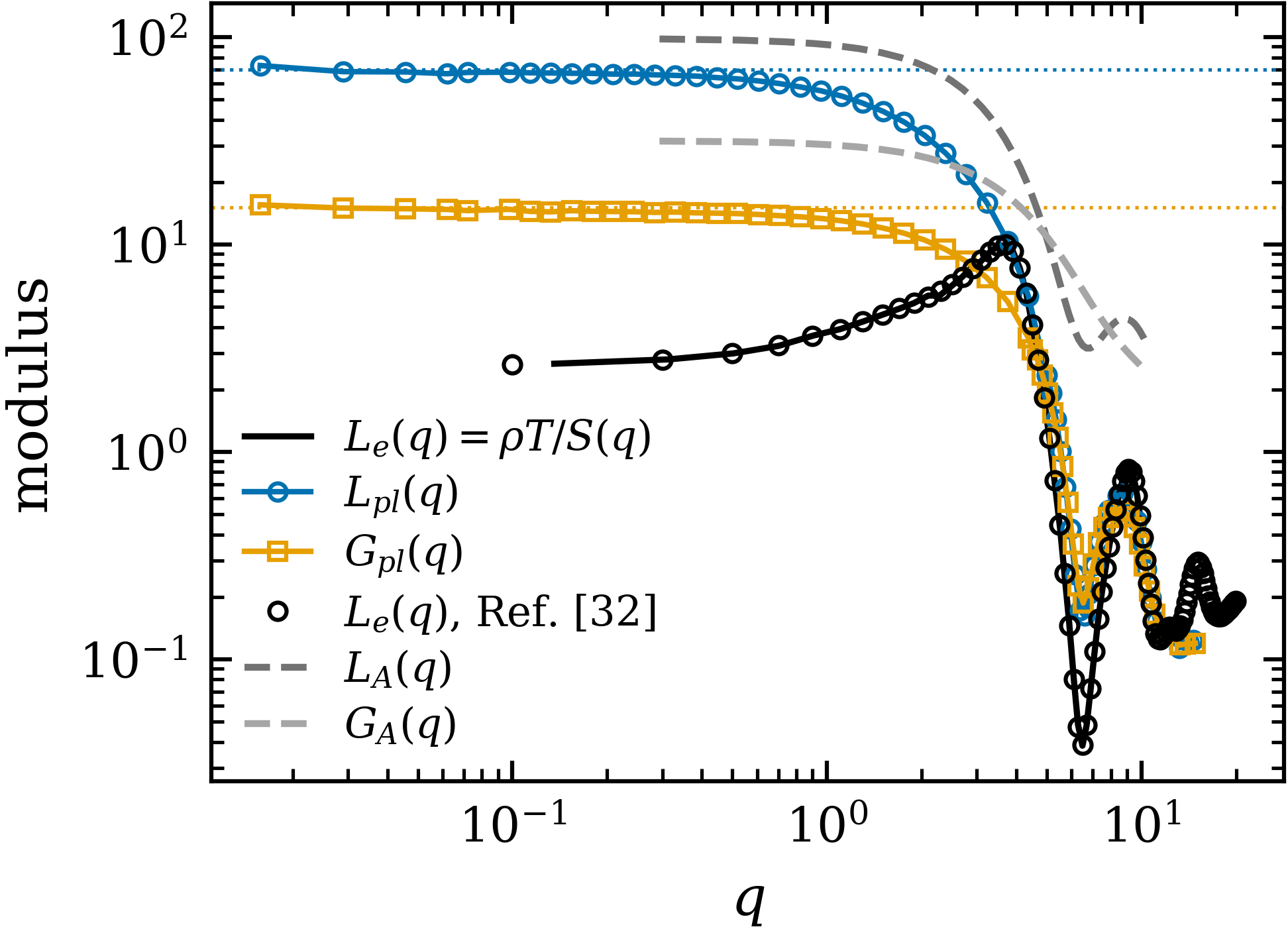}
    \caption{Plateau moduli $L_\mathrm{pl}(q)$ and $G_\mathrm{pl}(q)$ from
    Ref.~\cite{wittmer2024} ($N=160000$, averaged over all grid vectors
    with $|\qvec|\in[q-\delta q/2,\,q+\delta q/2]$) and
    $L_\mathrm{e}(q)=\rho T/S(q)$ from the structure factor of the present system
    at $T=0.2$. The dotted lines mark the $q\to0$ values of the plateau
    moduli. Dashed grey lines: affine moduli $L_A(q)$ and $G_A(q)$ of the present system at $T=0.2$.}
    \label{fig:WBfig2}
\end{figure}

The shear modulus $G_A(q)$ shows a monotonic decrease as 
$q$ grows, and so does the mixed modulus $M_A(q)$, while the longitudinal 
modulus $L_A(q)$ decreases until the minimum at $q\sim q_\mathrm{max}\approx 2\pi /b$, 
showing a weak increase at $q > q_\mathrm{max}$.  Here $q_\mathrm{max}\approx 2\pi /b$ corresponds to the peak of the structure factor $S(q)$ and $b$ denotes the average particle diameter in the polydisperse system. This feature is reminiscent of 
the behavior of the plateau modulus, $L_\mathrm{pl}(q)=L(q,t>t_\mathrm{pl})$ for $t$ beyond 
the onset time, $t_\mathrm{pl}$, of the glassy plateau, which was elucidated for the same 2DpLJ 
model well below the glass transition~\cite{wittmer2024}.
Interestingly, in the latter case, both the longitudinal and shear glassy 
moduli, $L_\mathrm{pl}(q)$ and $G_\mathrm{pl}(q)$, show dramatic minima near $q_\mathrm{max}$ (see 
Fig.~\ref{fig:WBfig2} and Fig.~2 in Ref.~\cite{wittmer2024}). 

It is remarkable that the equilibrium 
longitudinal modulus, $L_\mathrm{e}(q)=\rho T/S(q)$, see Refs.~\cite{semenov2024,grimm2025,sm2026}, shows an 
even deeper minimum:  while $\min L_\mathrm{pl}(q)\approx 0.2$, $\min L_\mathrm{e}(q)\approx 
0.04$ is 
much lower. Qualitatively, this behavior was expected since the equilibrium modulus 
describes the ultimate equilibrated state where the stress 
fluctuations are completely relaxed.  

It is also notable that the normal modulus, $N_A(q)$, exhibits a 
pronounced maximum at $q\sim q_\mathrm{max}$ (see Fig.~\ref{fig:affine_moduli}).  Another interesting 
observation is that kinetic contributions to affine moduli, $L_A^{\rm id}$, $M_A^{\rm id}$ 
and $N_A^{\rm id}$, show peaks at $q\approx q_\mathrm{max}\approx 2\pi /b$ (see Fig.~\ref{fig:affine_moduli_id}), while the total 
affine modulus, $L_A(q)$, exhibits a minimum there (see Fig.~\ref{fig:affine_moduli}).  

\begin{figure}[t]
    \centering
    \includegraphics[width=0.9\linewidth]{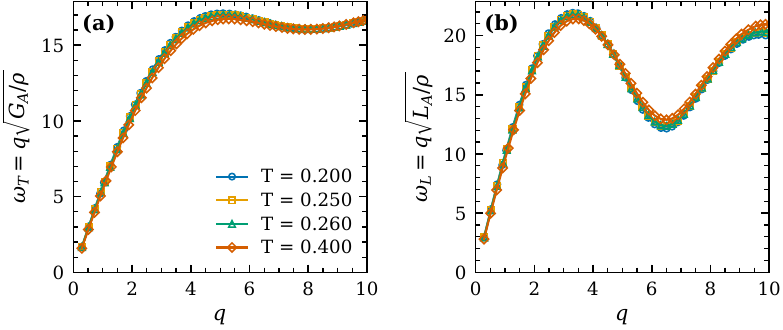}
    \caption{Upper-bound vibrational frequencies (a) $\omega_T=q\sqrt{G_A(q)/\rho}$
    and (b) $\omega_L=q\sqrt{L_A(q)/\rho}$ at the four temperatures.}
    \label{fig:newFomega}
\end{figure}

It is noteworthy that the elastic moduli $G_A(q)$ and $L_A(q)$ constitute the upper 
bounds of the relaxation moduli.  The upper-bound transverse and 
longitudinal sound velocities, therefore, are $v_T=\sqrt {G_A(q)/\rho}$ and 
$v_L=\sqrt {L_A(q)/\rho}$.  The corresponding vibrational frequencies, $\omega_T
=qv_T(q)$ 
and $\omega_L=qv_L(q)$, are shown in Fig.~\ref{fig:newFomega}. 

It is notable
that the frequency $\omega_L$ shows a maximum (followed by a minimum) as 
a function of $q$.  The maximum located at $q\approx q_\mathrm{max}/2$, and the 
minimum at $q\sim q_\mathrm{max}$ are rather pronounced.  Our idea is that these 
remarkable features (stemming from the $q$-dependent sound velocities) 
can be related to the so-called `boson peak'~\cite{SR2024,Zaccone2019} 
which is ubiquitous in amorphous solids.  Its nature is a subject of a 
long-lasting discussion~\cite{SR2024,Zaccone2019,GJ2023}. Interestingly, 
the frequency at the maximum gets somewhat higher at low 
temperatures, which is in line with the typical behavior of the boson 
peak frequencies~\cite{SR2024,GJ2023}. 

2. We have presented an algorithm and its implementation in LAMMPS for the direct
calculation of the wavevector-dependent microscopic stress tensor
$\sigma_{\alpha\beta}(\qvec,t)$ in molecular dynamics simulations of
amorphous materials. Note that  the  finite-$q$ stress
correlations and the corresponding generalized relaxation moduli have
been studied both theoretically and numerically in numerous previous
works~\cite{semenov2024,grimm2025,sm2026,lemaitre2014,lemaitre2015,klochko2018,lemaitre2017,maier2017,maier2018,wittmer2023,klochko2022}.  Many of these studies are based on the theoretical framework developed 
in~\cite{maier2017,klochko2018,maier2018,klochko2022,semenov2024} and the
subsequent extensions described in~\cite{grimm2025,sm2026}. The
contribution of the present work is to provide a general and portable
computational route from the pair forces generated by a molecular-dynamics
code to this wavevector-resolved stress field.

The central algorithm addresses the mismatch between the pairwise
structure of the molecular-dynamics force loop and the particle-wise
Fourier phase required by the stress tensor. Pair contributions are
accumulated in a particle-local buffer and the corresponding Fourier
phase is applied once per particle, using only the standard arguments of the
LAMMPS tally interface, without modifying the interaction implementation.
The result is a separation between the physical observable and the
particular pair potential used to generate the dynamics.

The second main feature is the use of an explicit list of wavevectors.
The wavevectors are therefore part of the input to the observable rather
than being fixed by an internally generated reciprocal-space grid. This
allows calculations to be targeted to the physically relevant
magnitudes and directions of $\qvec$, while avoiding the unnecessary
memory and computational cost associated with storing a complete
wavevector grid. The same interface can also be used to investigate
directional properties of tensor correlation functions without
confusing their tensorial coordinate-dependent components with intrinsic
anisotropy of the material~\cite{wittmer_tensor}.

Because a failure of a wavevector-resolved stress calculation can go unnoticed, validation is an essential part of the method. The
$\qvec=0$ limit provides an exact identity with the conventional stress
tensor calculated independently by LAMMPS and constitutes a particularly
strong test of the pairwise accumulation, kinetic contribution, and parallel
reduction. The finite-$q$ calculation is further tested against an
independent all-pairs evaluation, while the ideal contribution is checked
against analytical ensemble averages. Together, these tests provide
independent checks at the microscopic, numerical, and statistical theory levels.

The resulting stress field provides a direct connection between the variables obtained by 
microscopic simulations and the scale-dependent mechanical quantities (like $q$-dependent relaxation moduli)  whose relevance was highlighted in previous studies~\cite{semenov2024,klochko2018,klochko2022,grimm2025} . In particular, the zero-time stress correlations
provide estimators of the instantaneous, or affine, moduli, i.e.\ the four
affine moduli of Eq.~\eqref{eq:moduli}. These quantities represent the short-time mechanical response
before structural rearrangements can occur. Their subsequent time
dependence is continuously connected to the generalized relaxation moduli
$G(q,t)$, $K(q,t)$, and the associated tensorial response functions
introduced in~\cite{klochko2022,semenov2024}. The present implementation thus
provides the microscopic starting point required to study both the
instantaneous response and its relaxation across length and time
scales.

3. The computational framework also opens several directions for further
work. The same wavevector-resolved fields can be combined with density
and energy fluctuations to construct additional correlation channels
and mixed response functions. Extension to three-dimensional systems
requires only a few additional tensorial and vectorial components related to correlation functions and wavevectors,   while the accumulation scheme itself remains unchanged. At
small $q$, the time-dependent longitudinal and shear stress
correlations can be used furthermore to investigate acoustic propagation and
attenuation, while the connection between stress correlations and memory
kernels provides an important route to a more complete description of the viscoelastic
properties of  glass-forming and amorphous systems~\cite{grimm2025}.

More generally, the implementation described in the present paper has a potential to establish a reusable computational
interface between atomistic force calculations and spatially resolved
mechanical response. Since it uses only the standard tally interface, it works with
any pair style distributed with LAMMPS that is not a many-body potential, without
modifying it, which makes it applicable well beyond the simple liquid studied here.
Bonded interactions are not passed through the pair tally interface; for bead-spring
polymers their contribution would have to be added through an analogous bond callback.
Together with the arbitrary wavevector selection, this allows for a fast calculation of 
wavevector-resolved stress tensors  within the standard
molecular-dynamics workflow. The proposed approach therefore provides a numerical foundation for systematic studies of scale-dependent elasticity, stress
heterogeneity, long-range stress correlations, and viscoelastic
relaxation in amorphous materials.

\section*{Acknowledgements}

We acknowledge funding from the French National Agency for
Research ANR-24-CE30-7668 `VMFCMD', a grant of computer time at the HPC
computing cluster of the University of Strasbourg, and J. Wittmer for
providing initial 2D configurations. L. Klochko acknowledges support from the
CRET-AIREL, Universit\'e de Lorraine.

\section*{Conflict of interest}

The authors have no conflicts to disclose.

\section*{Data Availability}
\label{sec:avail}

The source of the computes, the correlator, the wavevector-list generator and
the reference implementation used in Sec.~\ref{sec:valid} are available at \url{https://github.com/liudmylaklochko/kstress}, under GPLv2. The computes are written for the LAMMPS version of
22 July 2025. 

\section*{Declaration of use of Generative AI}

Generative AI models were used for grammar and stylistic corrections/improvements of the text of this paper.

\appendix
\section{Implementation details}
\label{sec:impl_details}

\subsection{The ideal part in closed form}
\label{app:ideal}

For the kinetic contribution to the stress tensor given by the last term in Eq.~\eqref{eq:sigmamicro}  the ensemble averages of Eq.~\eqref{eq:moduli} (with the stress tensor replaced by its ideal kinetic term) 
can be evaluated exactly, because the velocities are Maxwellian and
independent of the positions and of each other. Writing
$u_j=\vec v_j\!\cdot\!\hat{\mathbf e}_1$ and $w_j=\vec v_j\!\cdot\!\hat{\mathbf e}_2$,
\begin{equation}
\sigma^{\rm id}_{11}(\qvec)=\frac{1}{V}\sum_j m\,u_j^2\,e^{\mathrm{i}\qvec\cdot\rvec_j},
\qquad
\sigma^{\rm id}_{12}(\qvec)=\frac{1}{V}\sum_j m\,u_j w_j\,e^{\mathrm{i}\qvec\cdot\rvec_j},
\end{equation}
and therefore
\begin{equation}
\big\langle|\sigma^{\rm id}_{11}|^2\big\rangle
=\frac{1}{V^2}\Big[\,\underbrace{N m^2\langle u^4\rangle}_{j=k}
 +\underbrace{m^2\langle u^2\rangle^2\!\sum_{j\neq k}\!\big\langle e^{\mathrm{i}\qvec\cdot(\rvec_j-\rvec_k)}\big\rangle}_{j\neq k}\Big].
\label{eq:app11}
\end{equation}

With $\langle u^2\rangle=T/m$, the Gaussian fourth moment
$\langle u^4\rangle=3\langle u^2\rangle^2$, and
$\sum_{j\neq k}\langle e^{\mathrm{i}\qvec\cdot(\rvec_j-\rvec_k)}\rangle=N S(q)-N$,
Eq.~\eqref{eq:app11} gives $N T^2[S(q)+2]/V^2$ and hence
$L_A^{\rm id}=\rho T\,[S(q)+2]$. The additive $2$ is thus $3-1$: the factor
three of the fourth moment of a single velocity component, minus the self term
removed in passing from $\sum_{j\neq k} (\ldots)$ to $S(q)$.

For $\sigma_{12}$ the two velocity components are different, so
$\langle u^2w^2\rangle=\langle u^2\rangle\langle w^2\rangle$ and
$\langle u w\rangle=0$, so that only the $j=k$ terms survive, giving
$G_A^{\rm id}=\rho T$. For the cross-correlation
$\langle\sigma_{11}\sigma_{22}^{*}\rangle$ the $j=k$ term is
$m^2\langle u^2w^2\rangle=T^2$ and cancels the self term removed from
$S(q)$, leaving $M_A^{\rm id}=\rho T\,S(q)$. Since $N_A$ differs from $L_A$
only by interchanging $\hat{\mathbf e}_1$ and $\hat{\mathbf e}_2$,
$N_A^{\rm id}=L_A^{\rm id}$  for any $\qvec$.

\subsection{Wavevectors: grid and list}
\label{sec:qlist}

The wavevectors for which $\sigma_{\alpha\beta}(\qvec)$ is evaluated are read
from a text file, one line per vector containing the two integers
$i_l\,j_m$, and the physical wavevector is
\begin{equation}
\qvec = 2\pi\left(\frac{i_l}{L_x},\ \frac{j_m}{L_y}\right),
\label{eq:qvec}
\end{equation}
where $L_x$ and $L_y$ are the box lengths along $x$ and $y$
($L_x=L_y=L$ for the square box used here), $i_l$ is an integer from
$-n_{q\max}$ to $n_{q\max}$, and $j_m$ an integer from $0$ to $n_{q\max}$.
The integer $n_{q\max}$ sets the largest wavevector,
$2\pi n_{q\max}/L$ along the axes; $n_{q\max}=120$ gives $q\approx7.3$ along
the axes and $q\approx10.3$ along the diagonal. Only the half plane
$j_m\geq0$ needs to be stored, since $\sigma(-\qvec)=\sigma^{*}(\qvec)$.

The list is arbitrary: it need not be a grid, need not have a common spacing,
and may mix signs. This matters for two reasons. The physically interesting
wavevectors are often a small and irregular subset of the full grid (a
few shells at small $q$ for the acoustic channels, and the single vector
$(0,0)$ for the validation of Sec.~\ref{sec:valid}), and the list allows them
to be addressed
directly. Then, the set of wavevectors becomes a property of
the run rather than of the source code so that changing it requires no
recompilation and no change of the analysis which reads the same file.

The earlier code used to obtain the results from Ref.~\cite{thesis,klochko2022}
generated the wavevectors internally as a full rectangular grid, writing one
line of the output file (a row) per $(i_l,j_m)$, the row index being
\begin{equation}
\text{row} = j_m + n_l\,(n_{q\max}+1),\qquad
i_l=n_l-n_{q\max}\in[-n_{q\max},n_{q\max}],\quad j_m\in[0,n_{q\max}].
\label{eq:gridrow}
\end{equation}
Supplying the file that reproduces Eq.~\eqref{eq:gridrow} gives the same
output, which is how the two were compared during development
(Sec.~\ref{sec:valid}).

The output columns involve the Cartesian components
$\sigma^{\rm Re}_{xy},\sigma^{\rm Im}_{xy},
 \sigma^{\rm Re}_{xx},\sigma^{\rm Im}_{xx},
 \sigma^{\rm Re}_{yy},\sigma^{\rm Im}_{yy}$,
and the projections~\eqref{eq:proj} are formed in post-processing. Keeping the
Cartesian components rather than the projections preserves the $\qvec=0$
entry, for which no longitudinal direction exists and which carries the check
against the ordinary stress.

\paragraph{Non-cubic boxes.}
The projections~\eqref{eq:proj} require the direction of the \emph{physical}
wavevector, $\hat{\mathbf e}_1\propto(i_l/L_x,\,j_m/L_y)$, which coincides with
the direction of the integer pair $(i_l,j_m)$ only for $L_x=L_y$. The earlier
analysis code~\cite{thesis,klochko2022} used the integer direction and was
therefore restricted to cubic boxes, while the code presented here calculates
$\hat{\mathbf e}_1$ from the physical components and accepts arbitrary box
aspect ratios. All results reported below were obtained for cubic boxes, for
which the two agree exactly.

\subsection{Accumulation through the tally interface}
\label{sec:trigger}

The virial term of Eq.~\eqref{eq:sigmamicro} is accumulated through the tally
callback of the pair style (Algorithm~\ref{alg:virial}). Since the
factor $\operatorname{e}^{-\mathrm{i}\qvec\cdot\rvec_i}$ refers to particle $i$ only,
the contributions of its neighbours are first collected in the buffer
$\sgmtmp^{\,i}$ of Eq.~\eqref{eq:buffer}, and the factor is applied once per
particle. On the other hand, the kinetic term is a direct sum over particles and involves no
callback (Algorithm~\ref{alg:kinetic}). Only the standard arguments of the
tally interface are used, so the compute works with the distributed pair
styles without requiring any further modification. The data flow is described in Sec.~\ref{sec:detail}.

\begin{algorithm}[t!]
\caption{Virial part of $\sgmout(\qvec)$ from the pair callback. Only the
standard \texttt{ev\_tally} arguments are used.}
\label{alg:virial}
\begin{algorithmic}[1]
\Require list of wavevectors $\{\qvec_k\}_{k=1}^{n_q}$ read from a file,
         sampling period $N$
\Ensure  the virial part of $\sgmout_{\alpha\beta}(\qvec_k)$ for every $k$, with
         $\sgmtmp$ the per-particle buffer of Eq.~\eqref{eq:buffer}
\State $\sgmout \gets 0$, \quad $\sgmtmp \gets 0$, \quad $i_{\rm cur} \gets -1$
       \Comment{once per sampled step}

\Statex
\Procedure{PairCallback}{$i,j,\Fvec_{ij},\rvec_{ij}$}
  \If{$t \bmod N \neq 0$ \textbf{ or } $f_{ij}=0$} \State \Return \EndIf

  \If{$i \neq i_{\rm cur}$}                     \Comment{a new particle has started}
     \If{$i_{\rm cur} \geq 0$} \State \Call{FinalizeAtom}{$i_{\rm cur}$} \EndIf
     \State $i_{\rm cur} \gets i$
  \EndIf

  \State $v_{\alpha\beta} \gets r_{ij,\alpha}\,F_{ij,\beta}/V$
  \For{$k = 1$ \textbf{to} $n_q$}
     \State $x \gets \tfrac{1}{2}\,\qvec_k\!\cdot\!\rvec_{ij}$
     \State $T \gets \operatorname{sinc}(x)\operatorname{e}^{\mathrm{i}x}$
            \Comment{phase of the pair midpoint, $T\to1$ as $x\to0$}
     \State $\sgmtmp_{k,\alpha\beta} \gets \sgmtmp_{k,\alpha\beta} + v_{\alpha\beta}\,T$
  \EndFor
\EndProcedure

\Statex
\Procedure{FinalizeAtom}{$i$}
  \For{$k = 1$ \textbf{to} $n_q$}
     \State $\sgmout_{k,\alpha\beta} \gets \sgmout_{k,\alpha\beta}
            + \operatorname{e}^{\mathrm{i}\qvec_k\cdot\rvec_i}\,\sgmtmp_{k,\alpha\beta}$
     \State $\sgmtmp_{k,\alpha\beta} \gets 0$
  \EndFor
\EndProcedure

\Statex
\Procedure{EndOfStep}{}
  \If{$i_{\rm cur} \geq 0$} \State \Call{FinalizeAtom}{$i_{\rm cur}$} \EndIf
  \State $\sgmout \gets \textsc{MPI\_Allreduce}(\sgmout,\ \textsc{sum})$
\EndProcedure
\end{algorithmic}
\end{algorithm}

\begin{algorithm}[t!]
\caption{Ideal (kinetic) part of $\sgmout(\qvec)$. A direct sum over the
particles: no pair style and no callback are involved.}
\label{alg:kinetic}
\begin{algorithmic}[1]
\Require the same list $\{\qvec_k\}$, sampling period $N$
\Ensure  $\sgmout^{\rm id}_{\alpha\beta}(\qvec_k)$

\If{$t \bmod N \neq 0$} \State \Return \EndIf
\State $\sgmout^{\rm id} \gets 0$
\For{each local particle $i$}
   \For{$k = 1$ \textbf{to} $n_q$}
      \State $\sgmout^{\rm id}_{k,\alpha\beta} \gets \sgmout^{\rm id}_{k,\alpha\beta}
             + \dfrac{m_i v_{i\alpha} v_{i\beta}}{V}\, \operatorname{e}^{\mathrm{i}\qvec_k\cdot\rvec_i}$
   \EndFor
\EndFor
\State $\sgmout^{\rm id} \gets \textsc{MPI\_Allreduce}(\sgmout^{\rm id},\ \textsc{sum})$
\end{algorithmic}
\end{algorithm}

\subsection{The algorithm in detail}
\label{sec:detail}

Since the construction is the central point of this paper, we describe it at
the level of the data flow, so that it can be reproduced for any other pair
quantity carrying a per-particle phase.

\paragraph{State.}

Let $n_q$ denote the number of wavevectors in the list, i.e.\ the number of
lines of the wavevector file (Sec.~\ref{sec:qlist}). The compute holds two
complex arrays of size $n_q\times3$ (six real columns each): the per-particle buffer $\sgmtmp$ of Eq.~\eqref{eq:buffer} and the
output accumulator $\sgmout$. It also holds the identity $i_{\rm cur}$ of the
particle whose neighbours are currently being delivered, and a record of the
last timestep on which it reset itself. The number of entries does not depend on the number of
particles since the memory is set by the number of wavevectors alone, which is
what makes a list of $10^4$ wavevectors affordable.

\paragraph{One callback.}
For every interacting pair the pair style calls the compute with the local
indices $i,j$, the scalar $f_{\rm pair}=|\Fvec_{ij}|/|\rvec_{ij}|$ and the
separation $\rvec_{ij}$ as these are the standard arguments of the tally
interface and nothing else is used. The compute first checks whether the
timestep is a sampling step and returns otherwise, so that on the other
steps its cost is one integer comparison per pair. On a sampling step it adds
the contribution of the pair to the buffer of the current particle. The first
callback of a sampling step also zeroes the accumulator, so that no state
survives from one sampled step to the next.
Note that the particle pair $(i,j)$ and the index pair $(i_l,j_m)$ are
unrelated: the former labels two interacting particles, the latter the two
integer components of a wavevector, Eq.~\eqref{eq:qvec}.

\paragraph{The pair kernel.}
The contribution of a pair to the buffer at wavevector $\qvec_k$ is
$v_{\alpha\beta}\,T(x_k)$, with $v_{\alpha\beta}=r_{ij,\alpha}F_{ij,\beta}/V$,
$T(x)=\operatorname{sinc}(x)\operatorname{e}^{\mathrm{i}x}$ and
$x_k=\tfrac12\qvec_k\!\cdot\!\rvec_{ij}$. Rather than evaluating a
transcendental function at every wavevector, the compute uses the fact that
the wavevectors are integer combinations of the reciprocal box vectors.
With $k_{1x}=\pi\Delta x/L_x$ and $k_{2y}=\pi\Delta y/L_y$ one has
$x_k=i_l k_{1x}+j_m k_{2y}$, so it tabulates, once per pair,
$\cos(nk_{1x}),\sin(nk_{1x})$ for $n=0,\dots,i_l^{\max}$ and
$\cos(nk_{2y}),\sin(nk_{2y})$ for $n=0,\dots,j_m^{\max}$, where
$i_l^{\max},j_m^{\max}$ are the largest index components in the list. The
sine and cosine of every $x_k$ then follow from the addition theorems with
two multiplications and one addition each, and negative indices are handled by
the parity of the sine. The $\operatorname{sinc}$ is a division by $x_k$,
guarded at $|x_k|<10^{-8}$ where it is set to one, the branch taken by the
$\qvec=0$ entry.

\paragraph{The atom factor.}
When the contributions of all pairs of particle $i$ have been accumulated, the buffer is multiplied by
$\operatorname{e}^{-\mathrm{i}\qvec_k\cdot\rvec_i}$ for every $k$ and added to the
accumulator: one complex multiplication per wavevector, incurred once per
particle rather than once per pair. That is the point of the two-level
structure: the factor saved is the number of neighbors per particle, about
twelve in the dense liquid. The sine and cosine of the atom factor are the
only transcendental calls made per particle, those of the tables are the
only ones made per pair, and the loop over wavevectors makes none. Since the
neighbour list is traversed particle by particle, the compute recognises that
particle $i$ is complete when the next callback carries a different particle
index. The last particle on each processor is completed when the output is
requested, before the reduction.

\paragraph{Reduction.}
The accumulators of all processors are summed with \code{MPI\_Allreduce}.
Since each pair is delivered on exactly one processor, with Newton's third law
on (the LAMMPS default), no pair is counted twice and the result is
independent of the domain decomposition, which we verify in
Sec.~\ref{sec:valid} by comparing runs on different numbers of ranks.

\paragraph{Assumptions.}
Two properties of the tally interface are relied upon: that the callbacks
of a given particle arrive contiguously, which holds because the neighbour
list is traversed particle by particle, and that a pair is delivered once,
which holds for half neighbour lists with Newton's law on. A pair style using
a full list with Newton's law off would deliver each pair from both ends and
double the result, which the $\qvec=0$ check detects immediately. No
particular order of the particles or of the neighbours within a particle is
assumed.

\paragraph{The kinetic part.}
Algorithm~\ref{alg:kinetic} is a plain sum over local particles with no
callback, since it needs only positions and velocities. It uses the same
list and the same atom-factor convention, so its output is row-aligned with
the virial part and the two are added column by column.

\subsection{Cost}
\label{sec:cost}

The work per sampled timestep has three parts. For each pair, the tables
$\cos(n k_{1x}),\sin(n k_{1x})$ and $\cos(n k_{2y}),\sin(n k_{2y})$, with
$k_{1x}=\pi\Delta x/L_x$ and $k_{2y}=\pi\Delta y/L_y$, where $\Delta x$ and
$\Delta y$ are the components of the pair separation $\rvec_{ij}$ (minimum
image), are filled for every integer $n$ up to the largest index component
of the list, which costs $2(i_l^{\max}+j_m^{\max}+2)$ transcendental
evaluations.

The kernel is then assembled for each of the $n_q$ wavevectors through the addition theorems, with arithmetics only. Finally, the atom factor costs $2n_q$ transcendental
evaluations per particle, since it is applied once per particle rather than
once per pair. In total,
\begin{equation}
W \simeq \underbrace{2\,(i_l^{\max}+j_m^{\max}+2)\,N_{\rm pair}}_{\text{tables}}
  \;+\; \underbrace{n_q\,N_{\rm pair}}_{\text{kernel, arithmetic}}
  \;+\; \underbrace{2\,n_q\,N_{\rm atom}}_{\text{atom factor}} ,
\label{eq:cost}
\end{equation}
where $n_q$ is the length of the list, i.e. the number of wavevectors on
which $\sigma_{\alpha\beta}(\qvec)$ is evaluated. The dominant term is the
loop over wavevectors for every pair, so that the cost is set by the number
of wavevectors rather than by the largest one. For a grid with indices up to
$n_{q\max}$ taken with step $\Delta n$ (only every $\Delta n$-th value of
$i_l$ and $j_m$ is kept), $n_q$ depends on $n_{q\max}$ and $\Delta n$ only
through their ratio: $n_{q\max}=120$ with $\Delta n=4$ gives $n_q=1891$, as
many as $n_{q\max}=30$ with $\Delta n=1$. For $\Delta n>1$ the tables are
filled up to $n_{q\max}$ rather than only at the multiples of $\Delta n$,
which makes the first term larger than in the earlier
code~\cite{thesis,klochko2022}. For the production grid, $n_{q\max}=120$
with $\Delta n=4$, we measure no difference in run time, the second term
being the larger one. What the list changes is the ability to ask for fewer
wavevectors: the second and third terms then fall in proportion to $n_q$,
the first with the largest index requested, and the volume of output falls
in proportion to $n_q$, which in practice is the more restrictive of the two
conditions.

\section{Porting to a current version of the MD code}
\label{sec:port}

The calculations in~\cite{thesis,klochko2022} were performed with
LAMMPS~\cite{lammps} of 7 September 2016. The implementation described here
targets the stable release of 22 July 2025. 

\subsection{The pair style}

The original calculation~\cite{thesis,klochko2022} relied on a custom pair
style, \code{lj/cut/rad}, written for the polydisperse Lennard-Jones model in
which the interaction range of a pair is set by the sum of the particle radii,
$\sigma_{ij}=R_i+R_j$. Current LAMMPS provides this functionality natively as
\code{lj/cut/sphere}, which reads the per-particle diameter from
\code{atom\_style sphere} and applies the cutoff $s_{\rm cut}\sigma_{ij}$ per
pair. The custom pair style is therefore no longer needed, and the model can
be run entirely with distributed code.

Replacing the custom pair style, however, breaks the earlier $\sigma(\qvec)$ compute used in~\cite{thesis,klochko2022}.
The earlier compute recognised the completion of a particle from the position of the
pair within the neighbour list, information that the modified pair style passed
through the tally interface together with the length of the list, and also
required the callback to be invoked for every list entry, including those
beyond the interaction cutoff. A distributed pair style calls the tally
interface with its standard list of arguments, in which the arguments needed by
the earlier compute contain other quantities. Evaluated on those arguments, the completion criterion is
never satisfied on a processor holding any particles: the accumulated
contributions are never transferred to the output array, and the virial part
of $\sigma(\qvec)$ is identically zero. The present compute uses only arguments that every pair
style supplies (Sec.~\ref{sec:detail}).

\subsection{Neighbour list and cutoffs}

The polydisperse model has two distinct cutoffs: the interaction range
$s_{\rm cut}\sigma_{ij}$, which depends on the pair, and the neighbour-list
range $R_n=R_f+\Delta s$, where $R_f$ is the largest interaction range the
pair style declares and $\Delta s$ the skin. Safety requires
$R_f=s_{\rm cut}\sigma_{\max}$, with $\sigma_{\max}$ the largest possible pair
diameter. If $R_f$ is understated, the effective skin is smaller than the
declared one, and the criterion used to trigger a rebuild, a displacement
exceeding half the \emph{declared} skin, no longer guarantees that no
interacting pair has been missed. Increasing the checking frequency does not
close this gap but only the correct declaration of $R_f$ does.
The current \code{lj/cut/sphere} declares $R_f=s_{\rm cut}\sigma_{\max}$, and
we verify in Sec.~\ref{sec:valid} that the forces produced with two very
different skins agree with a direct all-pairs calculation to machine
precision, with no dangerous builds reported.

\section{Usage}
\label{sec:usage}

\subsection{Installation}

The computes are installed by copying the source files into the LAMMPS
\code{src/} directory and rebuilding, and no patch to the distributed code is
required:

\begin{lstlisting}[style=lmpstyle]
cp compute_kstress_virial.cpp   $LAMMPS/src/
cp compute_kstress_virial.h     $LAMMPS/src/
cp compute_kstress_kinetic.cpp  $LAMMPS/src/
cp compute_kstress_kinetic.h    $LAMMPS/src/
cd $LAMMPS/build && cmake --build . -j
\end{lstlisting}

\subsection{The wavevector file}

All computes read the same text file, one line per wavevector, containing the
two integers $i_l\,j_m$ of Eq.~\eqref{eq:qvec}. Blank lines and lines
beginning with \code{\#} are ignored, and row $k$ of every output corresponds
to line $k$ of the file. Any set may be given. The generator supplied with the
code produces regular grids, but a list written by hand is equally valid.

\begin{lstlisting}[style=lmpstyle]
# il   jm
   0    0     # q = 0: the check against compute pressure
   4    0
   0    4
  -4    4
  .......
 120  120
\end{lstlisting}

\subsection{Running}

The computes are declared like any other LAMMPS compute, with the wavevector
file and the sampling period as arguments, and are written out with
\code{fix ave/time} in \code{mode vector}. Because the two computes share
the same wavevector file, their outputs are row-aligned.

\begin{lstlisting}[style=lmpstyle]
units           lj
atom_style      sphere
dimension       2
pair_style      lj/cut/sphere 2.2449
read_data       config.data
pair_coeff      * * 1.0
pair_modify     shift yes mix arithmetic
neighbor        0.3 bin
neigh_modify    every 1 delay 0 check yes
fix             f1 all enforce2d
set             atom * mass 1.0
timestep        0.005

# sampling period N of Algorithm 3: sigma(q) is computed every ksp steps
variable        ksp equal 4000

# virial and kinetic parts of sigma(q); qlist.dat = wavevector file (Sec. C.2)
compute         ks  all kstress_vir qlist.dat ${ksp}
compute         kin all kstress_kin qlist.dat ${ksp}

# write both outputs every ksp steps, one row per wavevector
fix fe all ave/time 1 1 ${ksp} c_ks[*]  format "%20.11e" file kstress.virial.dat  mode vector
fix fk all ave/time 1 1 ${ksp} c_kin[*] format "%20.11e" file kstress.kinetic.dat mode vector

# thermostat; T = temperature, given on the command line (-var T ...)
fix f3 all nvt temp ${T} ${T} 0.0129
run 20000000
write_restart   stage1.restart
\end{lstlisting}

Invocation is standard:

\begin{lstlisting}[style=lmpstyle]
mpirun -np Ncores lmp -in production.lmp -var T 0.400
\end{lstlisting}

\subsection{Analysis}

The correlator is given the same wavevector file and the output files. The
keyword \code{parts} adds the ideal and virial contributions separately. The
analysis program is independent of LAMMPS:

\begin{lstlisting}[style=lmpstyle]
g++ -O2 -fopenmp -o cort correlation_kstress.cpp -lm

./cort 50 qlist=qlist.dat kin=kstress.kinetic.dat exc=kstress.virial.dat parts
\end{lstlisting}

It writes one file per $|\qvec|^2$ shell containing the lag time and the
channels of Algorithm~\ref{alg:channels}. The projections assume a square box, $L_x=L_y$.

\subsection{Checking a new system}

Before any production run, include the row \code{0 0} in the wavevector file
and compare the corresponding output with the stress computed internally by
LAMMPS, split into its two parts:

\begin{lstlisting}[style=lmpstyle]
compute pv all pressure thermo_temp virial   # compare with kstress.virial.dat
compute pk all pressure thermo_temp ke       # compare with kstress.kinetic.dat
run 0
\end{lstlisting}

The two must agree to machine precision, up to the sign of the atom-factor
convention and the ordering of the components: LAMMPS returns
$xx,yy,xy$, the compute returns $xy,xx,yy$. This single comparison
exercises the pair loop, the completion of each particle and the parallel
reduction, and costs no simulation time.

\bibliographystyle{unsrt}
\bibliography{kstress}

\end{document}